\documentclass[useAMS,showpacs,a4paper,superscriptaddress,nofootinbib,tightenlines,floatfix]{raa}
\usepackage{graphicx,times}             %for PS/EPS graphics inclusion, new
\usepackage{natbib}
\usepackage{amssymb,amsmath}
\usepackage{epsfig}
\usepackage{float}
\usepackage{placeins}
\usepackage{graphicx}
\usepackage{epsfig}
\usepackage{bm}% bold math
\usepackage{amsfonts}
\usepackage{amssymb}
\usepackage{times}
\usepackage{natbib}
\usepackage{color}
\usepackage{multirow}
\usepackage[paperwidth=10in, paperheight=15.0in]{geometry}

\begin{document}
%\title{Intra-day variability timescales of three Seyfert Galaxies measured with XMM-Newton}
\title{Intra-day variability of three Seyfert Galaxies measured with XMM-Newton}
\subtitle{}

%     \volnopage{$Vol.0 (20xx) No.0,000$}  
%     \setcounter{page}{1} 
              
          \author{
          Ashutosh Tripathi \inst{1}
          \and Paul J.\ Wiita \inst{2,3}
          \and Alok C. Gupta \inst{4,5}
          \and Minfeng Gu \inst{5}
          \and M. Liao \inst{5,6}}
          
          \institute{ 
          Center for Field Theory and Particle Physics and Department of Physics, Fudan University, 220 Handan Road, Shanghai 200433, China
          \and
          Department of Physics, The College of New Jersey, P.O.\ Box 7718, Ewing, NJ 08628-0718, USA
          \and
          Kavli Institute for Particle Physics and Cosmology, SLAC, 2575 Sand Hill Rd., Menlo Park, CA, 94025, USA
          \and
          Aryabhatta Research Institute of Observational Sciences (ARIES), Manora Peak, Nainital - 263 002, India
          \and 
          Key Laboratory for Research in Galaxies and Cosmology, Shanghai Astronomical Observatory, Chinese Academy of Sciences, 80 Nandan Road, Shanghai 200030, China
          \and 
          University of Chinese Academy of Science, 19A Yuquanlu, Beijing 100049, China   \\       
          \vs\no
   {\small Received~~2018 month day; accepted~~2019~~month day}}

\abstract{
We present and analyze the variability of three Seyfert galaxies on intra-day timescales. We have analyzed in a uniform manner the 38 longest ( $>30$ ks) observations made for NGC 4051, MCG$-06-30-15$ and NGC 4151 by {\it XMM-Newton} between 2000 and 2015. The nuclei were quite active during essentially all of these observations and the overall X-ray fluxes (0.3--10 keV) varied by an order of magnitude. Most of the observations do appear to show characteristic timescales, estimated through their auto-correlation functions, ranging between $\sim 2.9$ ks and $\sim 45.3$ ks. The hard (2--10 keV) and soft (0.3--02 keV) bands are very well correlated but consideration of their hardness ratios show that the sources typically soften during flares. We also provide new estimates of the central black hole masses for these three AGN that support the hypothesis that Narrow Line Seyfert 1 galaxies have relatively small central black holes.
 }

%\begin{keywords}

%\end{keywords}
\maketitle

\section{Introduction}

Narrow-line Seyfert 1 galaxies (NLSy1) share most  general properties of Seyfert type 1 galaxies, including 
both narrow and broad emission lines and significant core flux variability,  but their permitted emission 
lines have relatively small widths.  The full width at half maximum of the H$\beta$ lines in these galaxies 
are  $\lesssim 2000$ km s$^{-1}$ (Goodrich 1989) with the ratio O[III]/H$\beta <3$ (Boller et al.\ 1996; 
Veron-Cetty et al.\ 2001). Their spectra show  signatures of Fe II multiplets which anti-correlate with the 
O[III] emission and with the width of broad Balmer lines in their spectra.  These characteristics support  the conclusion that 
the broad line regions and the accretion disks around their central supermassive black holes (SMBHs) can be 
viewed without any blockage, as is the case in general for Seyfert type 1 galaxies (e.g.\ Komossa \& Xu 2007). 
X-ray spectra of these galaxies show more variability than those of AGNs with broader emission lines.  All of 
these properties are consistent with their being AGNs with relatively low SMBH masses and high accretion rates 
and thus high Eddington ratios. Characteristic timescales for variability in AGNs have been proposed to scale 
directly with the SMBH mass (e.g.\ Markowitz et al.\ 2003; Papadakis 2004).  However, the NLSy1s do not neatly fit 
into this framework and  this behavior could be explained by the dependence of timescales on another parameter, 
such as black hole spin or accretion rate (e.g.\ Ishibashi \& Courvoisier 2009; Zhou et al.\ 2015; Pan et al.\ 2016).

The X-ray flux variations of AGNs on various timescales have, of course, been investigated frequently 
(e.g.\ Edelson \& Nandra 1999; 
Gaur et al.\ 2010; Gonz{\'a}lez-Martin \& Vaughan 2012; Kalita et al.\ 2015; Gupta et al.\ 2016; Pandey
et al.\ 2017, and references therein). Most of the sources considered for those studies are Seyfert galaxies 
or blazars as they exhibit the most rapid fluctuations in their observed fluxes, but Seyferts have the 
advantage in that they are much closer and usually have higher count rates, despite the Doppler boosting 
and temporal interval enhancements from which blazars benefit. Typical time-scales have been found to be 
less than a day, which is commonly known as intra-day variabilty (IDV) (Wagner \& Witzel 1995). In the last 
decade, there have been some claims for the presence of quasi-periodic oscillations (QPOs) in AGN X-ray time series 
data on IDV timescales (e.g. Espaillat et al.\ 2008; Gierli{\'n}ski et al.\ 2008; Lachowicz et al.\ 2009; 
Pan et al.\ 2016; and references therein). Variabilities of AGNs on these IDV timescales are the most puzzling. The study of characteristic AGN timescales are important as they can give insight 
about the emission mechanisms and the innermost regions of AGNs. We, therefore, have started examining 
X-ray IDV in Seyfert galaxies. In the first paper of this project, we present here an analysis of X-ray IDV measured using {\it XMM-Newton} as exhibited by two NLSy1 galaxies, NGC 4051 and MCG$-06-30-15$, as well as that of  NGC 4151, a Seyfert
1.5 galaxy.

Our aim in this paper is to search for and characterize IDV in Seyfert galaxies. We selected these sources because they have been  examined for rapid X-ray variability in the past and have been shown to be active. For instance, Vaughan et al.\ (2011) analyzed 17 pointed observations of NGC 4051 taken during May 2001 to June 2009. They stacked the light curves of all observations together and then computed a single power spectral density (PSD); however, they didn't analyze the individual light curves of NGC 4051 to examine them for IDV. For MCG$-06-30-15$, McHardy et al.\ (2005) performed a PSD analysis of the {\it RXTE} long-term light curve of MCG$-06-30-15$ that was observed between 1996 to 2004. They also included an analysis of some observations taken by {\it XMM-Newton}, producing a combined PSD analysis of 2--10 keV {\it RXTE} and 4--10 keV {\it XMM-Newton} light curves. In these earlier works, the galaxies were studied for the purpose of understanding the emission mechanisms and the accretion disk. Our studies, however, focus on searching for the variability of the order of a day, which meant we only analyzed the observations longer than 30 ks.  We are also concerned with any differences in the variability over soft and hard X-ray energy bands and so we also examined the cross-correlations between soft and hard X-ray bands. We also looked  at the correlation functions
  in order to search for any characteristic timescales or periodicities that might be present in the data.
  
 In Section 2 we discuss how we performed the data selection and reduction. In Section 3 we describe the  approaches 
we used to compute auto-correlation functions, excess variances, hardness ratios and cross-correlation functions. 
We give the results of those analyses in Section 4. A discussion and our conclusions are in Section 5.

\section{Data Selection and Reduction}
The aim of this work is to search for and characterize the X-ray intra-day variability of Seyfert galaxies using {\it XMM-Newton}. 
For this purpose, we looked into archival data from this satellite for Seyfert galaxies and chose 
 here those sources which have the most number of observations lasting longer than 30 ks, which are among 
 the sources most often observed by {\it XMM-Newton}.
Thanks to its excellent time resolution and ability to  continuously point at specific X-ray 
sources for extended times, {\it XMM-Newton} is a superb X-ray observatory for time series 
analyses (Jansen et al.\ 2001). We have selected and analyzed time series data taken during
2000 to 2015 of three Seyfert galaxies: a total of 17 observations 
of NGC 4051,  7 of MCG$-06-30-15$, and 14 of NGC 4151 were taken 
when the sources were bright enough to provide sufficient count rates.
We downloaded the data  
from the {\it XMM-Newton} Science archive.
Our key selection criterion was that each observational 
duration exceeded 8 hours so we could probe 
and characterize the intra-day variability (IDV) 
from tens of minutes to a few hours. These observations use the {\it European Photon Imaging Camera 
(EPIC)} which has the ability to track the source continuously for many hours and its wide field of 
view  is extremely beneficial for background subtraction. EPIC consists of three co-aligned CCD 
telescopes, MOS1, MOS2 and pn (Turner et al.\ 2001; Str{\"u}der et al.\ 2001). We only use the EPIC/pn 
(and not EPIC/MOS) because the former is least affected by pileup effects. We choose the 
energy range of 0.3$-$10.0 keV for the light curve (LC) extraction because the data below 0.3 keV are significantly
affected by pileup while those above 10 keV they are seriously affected by background 
noise.  All the observations considered here were taken in the small window (SW) mode.

We used {\it XMM-Newton} Science Analysis Software (SAS) version 15.0.0 for the data reduction.
Event files were generated using the standard SAS routine \textit{epchain}. The good time intervals 
(GTIs) were generated using \textit{TABTIGEN} which contains the data devoid of proton flares. Then 
the cleaned event files were generated by combining the GTI and event list files and filtered using 
single and double event files (PATTERN $\leq$ 4) but excluding the events that are at the edge of the 
CCD (FLAG=0). To obtain the source event files, we have extracted an area of radius 30 -- 40 arcsec, depending 
on the source location on the CCD. For background events, an area of radius 40 -- 50 arcsec, located as far as 
away as possible from the source, is chosen to avoid any of its contribution. Using \textit{epiclccorr}, 
the background subtracted LCs were generated for the time bin of 100 seconds. We have checked for pileup 
effects for these observations using the SAS routine \textit{epatplot} and found no significant pileup
in any of the observations studied here. 

Table 1 lists the observation log of {\it XXM-Newton} data for these three Seyfert galaxies. 
The first column provides the observation ID and the second column, the date when the observation was taken. The 
third column gives the revolution number of the satellite when the observation was taken. The fourth and 
fifth columns list the total good exposure time (GTI) and the mean count rate, $\mu$, respectively.  

\begin{table}
{\bf Table 1.} Log of XMM-Newton observations of three NLSy1s. \\
\centering
\scriptsize
\noindent
\begin{tabular}{cccrr}\hline \hline
Observation &Date of      &Revolution &GTI      & $\mu$ (ct  s$^{-1}$) \\ 
ID          & Observation &           &(sec)    & \\\hline
\multicolumn{5}{|c|}{NGC 4051}\\ \hline
0109141401 & 2001.05.16 & ~263 & 116513 & 28.87$\pm$0.02 \\
0157560101 & 2002.11.22 & ~541 &  49710 &  5.74$\pm$0.12\\
0606320101 & 2009.05.03 & 1721 &  45236 & 11.49$\pm$0.02 \\
0606320201 & 2009.05.05 & 1722 &  44346 & 18.02$\pm$0.02 \\
0606320301 & 2009.05.09 & 1724 &  31268 & 22.27$\pm$0.03 \\
0606320401 & 2009.05.11 & 1725 &  28750 &  5.10$\pm$0.02 \\
0606321301 & 2009.05.15 & 1727 &  30138 & 25.98$\pm$0.04 \\
0606321401 & 2009.05.17 & 1728 &  37992 & 16.42$\pm$0.03 \\
0606321501 & 2009.05.19 & 1729 &  38789 & 16.02$\pm$0.02 \\
0606321601 & 2009.05.21 & 1730 &  41453 & 33.49$\pm$0.03 \\
0606321701 & 2009.05.27 & 1733 &  38297 &  7.20$\pm$0.02 \\
0606321801 & 2009.05.29 & 1734 &  39886 & 11.61$\pm$0.02 \\
0606321901 & 2009.06.02 & 1736 &  36419 &  5.75$\pm$0.01 \\
0606322001 & 2009.06.04 & 1737 &  36893 & 10.97$\pm$0.02 \\
0606322101 & 2009.06.08 & 1739 &  37621 &  3.32$\pm$0.01 \\ 
0606322201 & 2009.06.10 & 1740 &  41084 &  9.49$\pm$0.01 \\
0606322301 & 2009.06.16 & 1743 &  41978 & 12.23$\pm$0.02 \\ \hline
\multicolumn{5}{|c|}{MCG$-06-30-15$}\\ \hline
0111570101 & 2000.07.11 & ~108  &  41104 & 15.56$\pm$0.02 \\
0029740101 & 2001.07-31 & ~301  &  83335 & 27.71$\pm$0.02 \\
0029740701 & 2001.08-02 & ~302  &  126169 & 29.71$\pm$0.02 \\
0029740801 & 2001.08-04 & ~303  &  124369 & 27.62$\pm$0.02 \\
0693781201 & 2013.01-29 & 2407  & 133710  & 36.43$\pm$0.02 \\ 
0693781301 & 2013.01-31 & 2408  & 133640  & 21.49$\pm$0.02 \\
0693781401 & 2013.02-02 & 2409  &  48459  & 27.71$\pm$0.02 \\ \hline
\multicolumn{5}{|c|}{NGC 4151}\\ \hline
0112310101 & 2000.12.21 & ~190  &   29963 & 7.13$\pm$0.02\\
0112830201 & 2000.12.22 & ~190  &   56988 & 7.44$\pm$0.02\\
0112830501 & 2000.12.22 & ~190  &   19700 & 7.41$\pm$0.06\\
0402660101 & 2006.05.16 & 1178 &   39932 & 6.42$\pm$0.02\\
0402660201 & 2006.11.29 & 1277 &   45611 & 10.43$\pm$0.04\\
0761670101 & 2015.11.12 & 2917 &   33574 & 7.39$\pm$0.02\\
0761670201 & 2015.11.14 & 2918 &   41116 & 8.57$\pm$0.04\\
0761670301 & 2015.11.16 & 2919 &   43654 & 7.41$\pm$0.03\\
0761670401 & 2015.11.30 & 2926 &   43443 & 7.73$\pm$0.03\\
0761670501 & 2015.11.20 & 2921 &   38345 & 9.56$\pm$0.02\\
0761670601 & 2015.11.22 & 2922 &   42612 & 9.86$\pm$0.03\\
0761670701 & 2015.12.16 & 2934 &   48023 & 12.48$\pm$0.02\\
0761670801 & 2015.12.18 & 2935 &   43435 & 13.49$\pm$0.04\\
0761670901 & 2015.12.22 & 2937 &   43942 & 11.93$\pm$0.02\\\hline
\end{tabular}
\end{table}
  
\section{Light Curve Analysis Techniques}

It is well established that many subclasses of AGNs, 
including Seyfert galaxies. show IDV in their X-ray LCs (e.g.\ Gaur et al.\ 2010; Gonz{\'a}lez-Martin \& Vaughan 2012, 
Kalita et al.\ 2015; Gupta et al.\ 2016; Pandey et al.\ 2017, and references therein). 

The {\it XMM-Newton} LCs of  NGC 4051, MCG$-06-30-15$ and NGC 4151 for the observation IDs
given in Table 1 are plotted in Figure 1. We made these LCs in the broad energy range 0.3--10 keV. By visual 
inspection, we can conclude  that IDV is present in almost all the LCs and that during most of them significant 
flares were measured that rose and fell over the course of an hour or so. To obtain  quantitative estimates of 
IDV timescales,  
if present, we have   computed auto-correlation function (ACF)
of the LCs to search for characteristic timescales in the IDV. 
Excess variance and fractional variance are used to calculate the IDV parameters. 
The X-ray variability parameters of observations are reported in Table 2. Brief descriptions of how we computed ACFs 
and excess variances are given in following subsections. 

\subsection{Auto-Correlation Function}

The auto-correlation function (ACF) is the correlation between two values of the same variable at discretized 
times $t_i$ and $t_{i+k}$. The ACF shows  the relation between the past and future points of the data 
separated by a given time interval and so indicates the randomness or periodicity of the data.
Given the measurements $X_1$, $X_2$,...$X_N$, with mean $\bar{X}$, measured at times 
$t_1$, $t_2$,..$t_N$, the ACF ($r_k$) at lag $k$ for equi-spaced observations is defined as

\begin{equation}
r_k = \frac{\Sigma_{i=1}^{N-k}(X_{i}-\bar{X})(X_{i+k}-\bar{X})}{\Sigma_{i=1}^{N}(X_{i}-\bar{X})^2}.
\end{equation}

An ACF measures the relative strength of the signal and noise, as the presence of noise in the signal causes 
the value of ACF to decrease with time.  
The presence of a peak at other than $\tau$ = 0 signals the possible presence of periodicity of a QPO but the dips 
indicate the presence of characteristic timescales in the data (see section 3.2 of Pandey et al.\ 2017). 
If the binning is  appropriate, the computed ACF will appear as the even function it should be, having
the same value at the time lags $t$ and $-t$ (Hufnagel \& Bregman 1992; Hovatta et al.\ 2007). 

In every case for these observations of NGC 4051, MCG$-06-30-15$ and NGC 4151, the ACFs gave putative timescales 
where minima are seen and where the monotonic declines away from the peak at $t = 0$ are followed by (usually 
modest) rises (Pandey et al. 2017); these dip timescales are listed as $\tau$ in Table 2 and range from 2.9 ks 
through 45.3 ks. 
In no case were there repetitive dips at nearly the same  multiples of a lag time that might signal a QPO.

\subsection{Excess Variance}

Actual light curves are comprised of $N$ flux measurements, $x_i$, along with their uncertainties, $\sigma_i$, 
for the $i^{th}$ time bin. The error arises mainly from measurement errors but as the typical X-ray source yields 
 rather few photons per second, the error may also have a substantial Poisson contribution that adds  to the total 
uncertainty. In order to obtain the intrinsic source variance, these errors needed to be subtracted (Nandra 
et al.\ 1997; Edelson et al.\ 2002). Let the mean square error, ${\overline{\sigma_{err}^2}}$, and sample variance, $S^2$, 
be defined respectively as

\begin{equation}
{\overline{\sigma_{err}^2} }= \frac{1}{N}\Sigma_{i=1}^N \sigma_i ^2 ,
\end{equation}
and 
\begin{equation}
S^2  = \frac{1}{N-1}\Sigma_{i=1}^N(x_i-\bar{x})^2.
\end{equation}
Then the excess variance is defined as
\begin{equation}
\sigma_{XS}^2 = S^2 - {\overline{\sigma_{err}^2}}.
\end{equation}

The normalized excess variance is found by dividing the excess variance by the mean of the flux and the fractional root mean square (rms) variability is defined as its square root (Edelson et al.\ 1990; Rodr{\'{\i}}guez-Pascual et al.\ 1997):
\begin{equation}
F_{var} = \sqrt{\frac{ S^2 - {\overline{\sigma_{err}^2}}} {\bar{x}^2}}.
\end{equation}
This $F_{var}$ gives the average variability amplitude with respect to the mean flux of the source and thus provides the 
rms variability in fractional terms. The uncertainty in the rms variability was calculated by Monte-Carlo 
simulations described in Vaughan et al.\ (2003) and can be given by (Jaffe et al.\ 2004) 
\begin{equation}
(F_{var})_{err}=\sqrt{ \left\{ \sqrt{\frac{1}{2n}} \frac{ \overline{\sigma_{err}^{2}}
}{\bar{x}^{2}F_{var} } \right\}^{2}+\left\{ \sqrt{\frac{\overline{\sigma_{err}^{2}}}{n}}\frac{1}{\bar{x}}\right\}^{2}}.
\end{equation}

Table 2 contains X-ray variability parameters for NGC 4051, MCG$-06-30-15$ and NGC 4151. 
The second and third columns represent excess variance and normalized excess variance, respectively, for the entire X-ray 
band.  $F_{var}$ values are given, along with their 
errors, in the fourth through sixth columns for the entire X-ray band considered, as well as for fluxes divided into 
soft (0.3--2.0 keV) and hard (2.0-10.0 keV) bands, respectively. The seventh %and eighth 
column of Table 2 gives  
the timescales indicated by the ACF for the overall LCs. 

\subsection{Soft and hard X-ray band correlations}

The spectrum of the source can give information about the physical processes responsible for its 
variation (Park et al.\ 2006 and references therein). Study of the spectra of each of these 
observations is beyond the scope of this paper, and in many cases the modest count rates would 
preclude us from producing decent spectra.  However, there was 
a sufficiently high count rate in nearly every one of these {\it XMM-Newton} observations to allow
us to examine the variations in the soft and hard 
bands separately.  These individual LCs are shown in the left columns of Figure 3. It is clear from 
those figures that usually the fractional variations in the soft band exceed those in the hard, 
though there are exceptions.  This impression in quantified through values of $F_{var}$ computed 
and shown in the fourth through sixth columns of Table 2.

The cross correlation function (CCF) is the measure of the degree to which the data points of
different time series are correlated at a particular  time delay. The CCF calculates how the  points
of a particular data set ($X_1, X_2,...,X_i,...,X_N$) are related to the points of another data
set ($Y_1, Y_2,..., Y_{i+k}, ...,Y_N$),  occurring with a time delay $k$ as,
\begin{equation}
r_k = \frac{\Sigma_{i=1}^{N-k}(X_{i}-\bar{X})(Y_{i+k}-\bar{Y})}{\sqrt{\Sigma_{i=1}^{N}(X_{i}-\bar{X})^2}
\sqrt{\Sigma_{i=1}^{N}(Y_{i}-\bar{Y})^2}}.
\end{equation}
Here  $\bar{X}$ and $\bar{Y}$ are the means of the $X$ and $Y$ data, respectively. The CCFs between the 
soft and hard X-ray bands are given in the middle columns in Figure 3.  A peak at a non-zero time indicates 
that one of the  bands consistently led the other as they varied.  When the two data sets are same, then the 
CCF is simply the ACF discussed in Section 3.1. 

As is frequently done, we employ the hardness ratio (HR) as a simple proxy for the spectrum, defining it as 
\begin{equation}
HR = \frac{H-S}{H+S},
\end{equation}
where here $H$ refers to the hard X-ray flux (2.0--10 keV) and $S$ is its soft counterpart
(0.3--2.0 keV). These HR plots as functions of time are shown in the right columns of Figure 3. 

\section{Results}

\subsection{NGC 4051}

NGC 4051 is a NLSy1 galaxy first observed in X-rays by the {\it Einstein Observatory} (Marshall et al.\ 1983). 
Peterson et al.\ (2000) claimed that the frequently seen rapid X-ray variability  is not observed in optical observations 
even when correlated on the timescales of many weeks and longer.  Lamer et al.\ (2003) reported the X-ray 
variability observed by {\it RXTE} observations during 1996--1999 during which the flux of NGC 4051 varied 
by a factor of about 100.

\begin{subfigures}
\begin{figure*}[t]
\centering
%\vspace*{-0.2in}
\vspace*{-0.1in}\epsfig{figure= 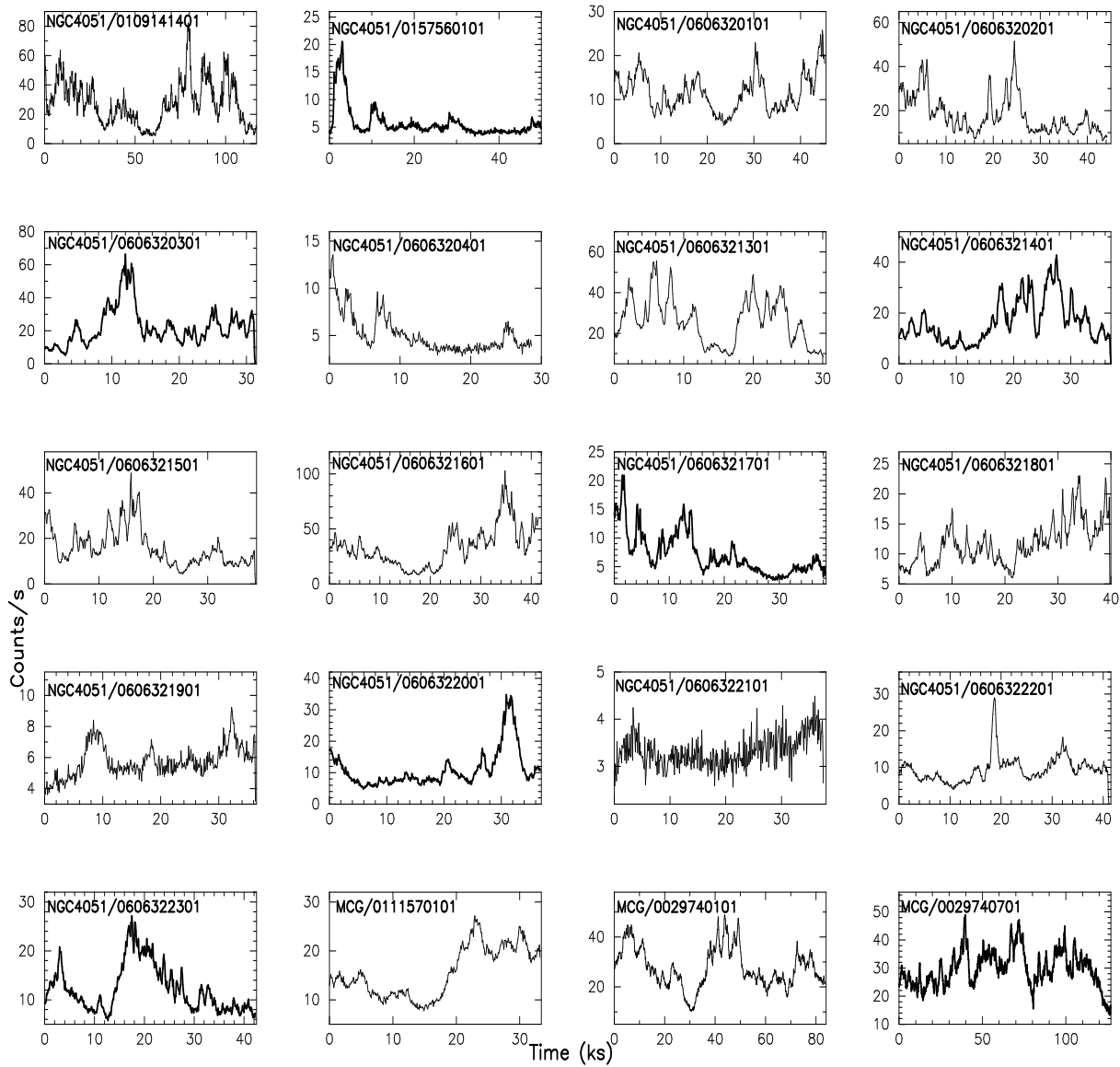,height=6.0in,width=6.0in,angle=0}
\caption{XMM-Newton light curves of the Seyfert galaxies NGC 4051 and  MCG$-06-30-15$ (MCG) with the name and 
observation ID written in each panel.}
\label{}
\end{figure*}

\begin{figure*}
\centering
%\vspace*{-0.2in}
\vspace*{-0.1in}\epsfig{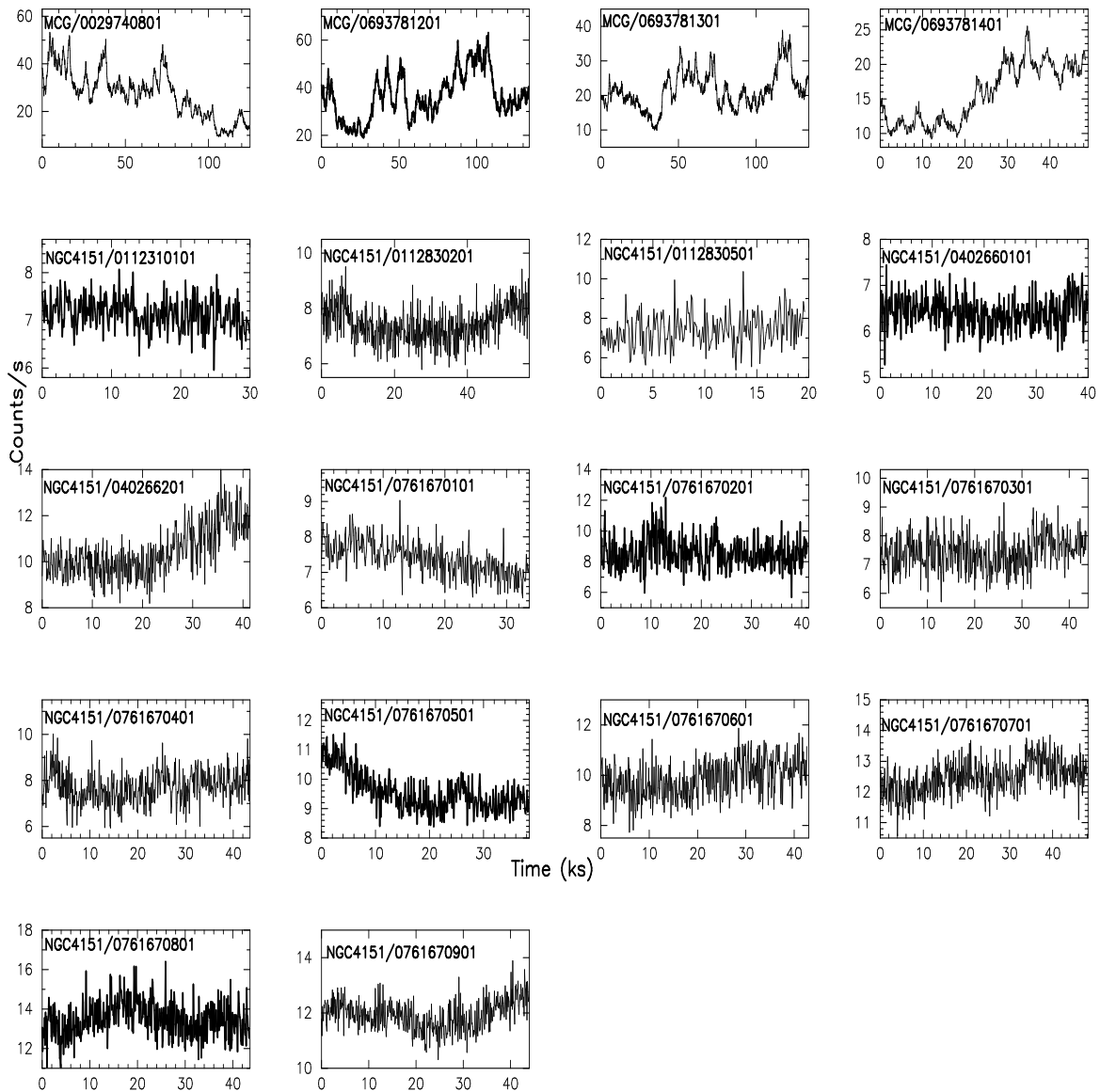}
\caption{As in Fig.\ 1a for MCG$-06-30-15$ (MCG) and NGC 4151.}
\label{}
\end{figure*}
\newpage
\end{subfigures}

\begin{subfigures}
\begin{figure*}[t]
\centering
%\vspace*{-0.2in}
\epsfig{figure= 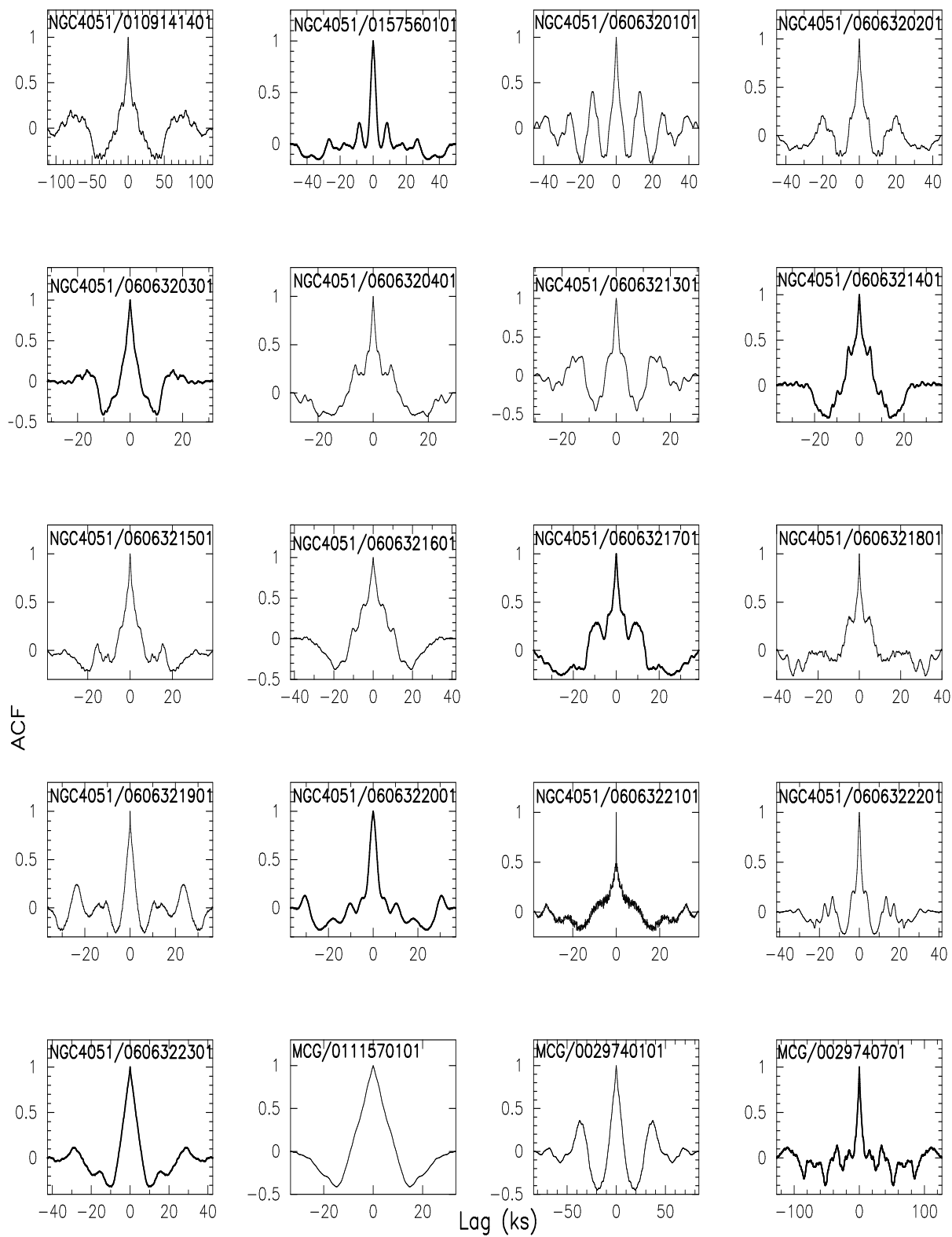,height=8.0in,width=6.0in,angle=0}
\caption{Auto-correlation functions (ACFs) for NGC4051 and MCG$-06-30-15$ (MCG), with object name and observation number labeling each panel. }
\label{}
\end{figure*}

\begin{figure*}
\centering
%\vspace*{-0.2in}
\epsfig{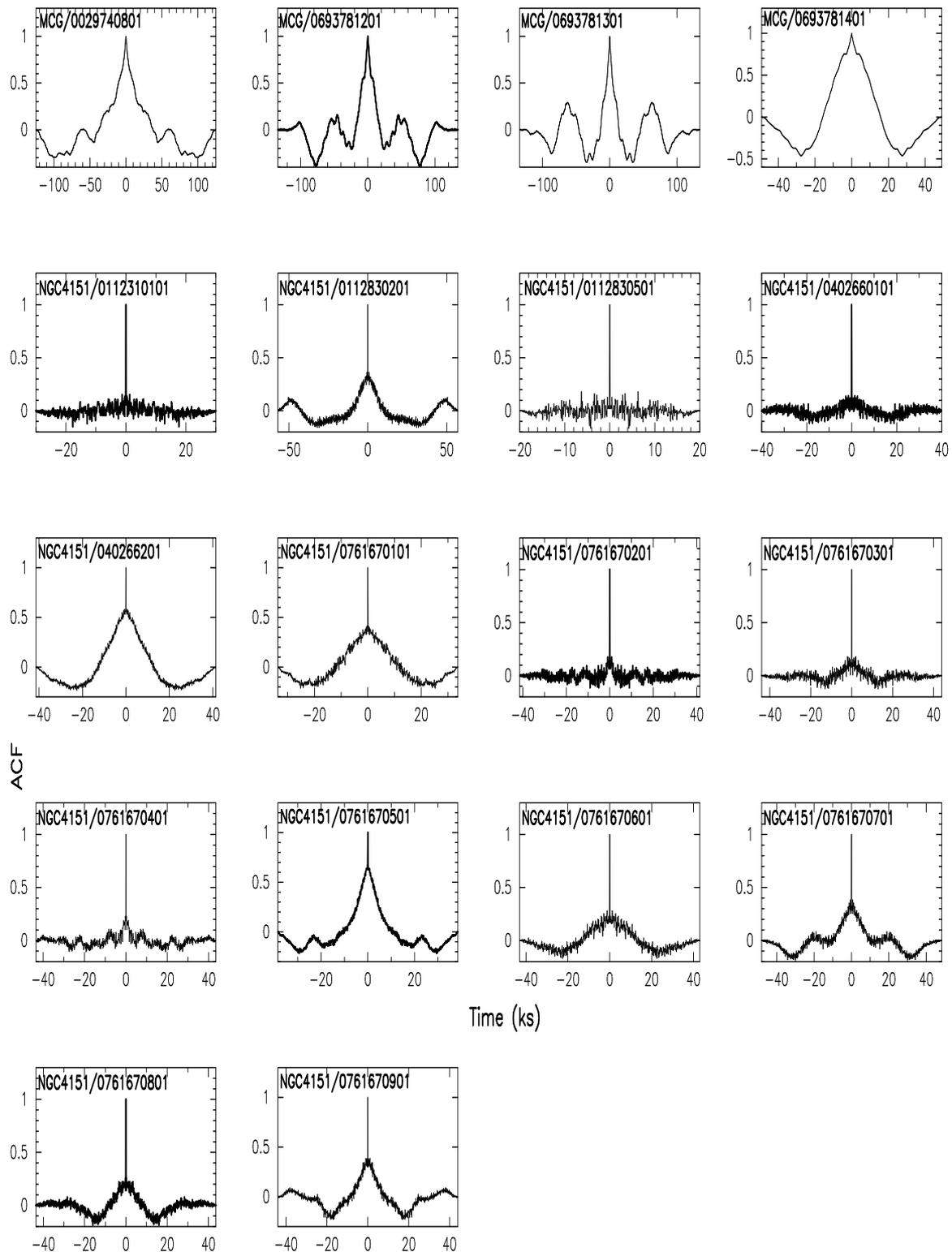}
\caption{ACFs, as in Fig.\ 2b for MCG$-06-30-15$ (MCG) and NGC 4151. }
\label{}
\end{figure*}
%\newpage
\end{subfigures}

\begin{table*}
\centering
{\bf Table 2.} X-ray Variability Parameters for NLSy1s. \\
\noindent
\scriptsize
\begin{tabular}{crcccccc}\hline \hline
Observation &$\sigma_{XS}^2$&$\sigma_{NXS}^2$& &$F_{var}$(percent)& &$\tau$\\ 
ID        &  (total)    & (total)   &   Total(0.3-10keV)&Soft(0.3-2keV)&Hard(2-10keV)   & (ks)   \\
\hline
\multicolumn{7}{|c|}{NGC 4051}\\ \hline
0109141401 & 216.64 & 0.28 & 52.82$\pm$0.07&54.88$\pm$0.08&37.93$\pm$0.23 &{45.3}\\
0157560101 & 8.11   & 0.24 & 49.37$\pm$0.21&49.23$\pm$0.22&57.92$\pm$0.58 &  ~4.7\\
0606320101 & 17.14  & 0.13 & 36.01$\pm$0.18&38.7$\pm$0.2&26.64$\pm$0.46   & ~5.6 \\
0606320201 & 74.53  & 0.23 & 47.92$\pm$0.14&50.04$\pm$0.15&35.44$\pm$0.42 &{~9.3} \\ 
0606320301 & 133.25 & 0.27 & 51.90$\pm$0.15&53.83$\pm$0.16&39.73$\pm$0.48 &{10.1} \\ 
0606320401 & 4.06   & 0.16 & 39.48$\pm$0.36&40.97$\pm$0.39&37.66$\pm$0.86 &{~3.5} \\
0606321301 & 126.8  & 0.19 & 43.41$\pm$0.14&43.96$\pm$0.15&39.39$\pm$0.51 &{~7.6} \\
0606321401 & 69.39  & 0.26 & 50.89$\pm$0.16 &52.72$\pm$0.17&39.41$\pm$0.52&{13.7} \\
0606321501 & 67.05  & 0.26 & 51.19$\pm$0.16 &53.64$\pm$0.17&36.28$\pm$0.51&{12.7}\\
0606321601 & 315.81 & 0.28 & 53.06$\pm$0.10&54.66$\pm$0.11&40.59$\pm$0.34 &{19.1} \\
0606321701 & 13.42  & 0.26 & 50.94$\pm$0.25&53.84$\pm$0.27&41.32$\pm$0.61 &{~5.7}\\
0606321801 & 12.18  & 0.09 & 30.11$\pm$0.20&30.41$\pm$0.21&29.08$\pm$0.5 & ~2.9 \\
0606321901 & 0.99   & 0.03 & 16.68$\pm$0.29&15.2$\pm$0.32&29.99$\pm$0.82 & ~6.4 \\ 
0606322001 & 35.32  & 0.03 & 54.25$\pm$0.21&56.39$\pm$0.22&46.36$\pm$0.59 &{~7.6} \\ 
0606322101 & 0.08   & 0.01 & ~6.81$\pm$0.49&7.89$\pm$0.49&26.92$\pm$1.47 &{17.6} \\ 
0606322201 & 12.73  & 0.14 & 37.64$\pm$0.21&39.31$\pm$0.23&34.51$\pm$0.58 &{~8.3}\\
0606322301 & 22.46  & 0.15 & 38.69$\pm$0.18&40.48$\pm$0.19&28.54$\pm$0.56 & ~9.6 \\ \hline
\multicolumn{7}{|c|}{MCG-06-30-15}\\ \hline
0111570101 & 25.92 & 0.11 & 32.73$\pm$0.18&34.68$\pm$0.2&26.32$\pm$0.39  & 14.8 \\
0029740101 & 63.54  & 0.08 & 28.76$\pm$0.08&29.78$\pm$0.09&25.59$\pm$0.2  &{19.4} \\
0029740701 & 39.75  & 0.04 & 21.23$\pm$0.07&21.66$\pm$0.07&20.36$\pm$0.15  & 10.6\\
0029740801 & 96.32 & 0.13 & 35.53$\pm$0.07&36.13$\pm$0.08&33.86$\pm$0.16  & {44.0} \\
0693781201 & 91.16 & 0.07 & 26.2$\pm$0.06&26.39$\pm$0.06&26.32$\pm$0.13  & {24.0} \\
0693781301 & 28.87 & 0.06 & 24.99$\pm$0.08&25.01$\pm$0.08&26.04$\pm$0.17  & {34.6} \\
0693781401 & 17.37 & 0.07 & 26.53$\pm$0.15&27.53$\pm$0.17&23.6$\pm$0.32  & {27.4} \\ \hline
\multicolumn{7}{|c|}{NGC 4151}\\ \hline
0112310101 & -  & - & -0.38$\pm$0.24&-&1.45$\pm$1.17  &-\\
0112830201 & 0.14 & 0.002 & 4.97$\pm$0.4&3.82$\pm$0.98&8.04$\pm$0.5  & {30.8} \\
0112830501 & 0.03 & 0.01 & 2.45$\pm$2.87&6.96$\pm$2.69&- &-\\
0402660101 & 0.01 & 0.0002 & 1.69$\pm$0.6&1.52$\pm$1.48&22.7$\pm$0.78  & { 18.8} \\
0402660201 & 0.62 & 0.005 & 7.55$\pm$0.4&-&9.74$\pm$0.47  &22.4\\
0761670101 & 0.1 & 0.002 & 4.26$\pm$0.34&-&7.04$\pm$0.41  &20.5\\
0761670201 & 0.23 & 0.003 & 5.56$\pm$0.82&-2.09$\pm$4.34&5.47$\pm$1.26  &-\\
0761670301 & 0.02 & 0.001 & 2.28$\pm$0.98&2.52$\pm$1.958&4.67$\pm$0.92 &-\\
0761670401 & 0.02 & 0.002 & 3.97$\pm$1.39&3.97$\pm$1.398&5.97$\pm$0.95  &-  \\
0761670501 & 0.28 & 0.003 & 5.57$\pm$0.24&-&8.8$\pm$0.28  & {21.4} \\
0761670601 & 0.1 & 0.001 & 3.17$\pm$0.6&-&5.28$\pm$0.61  & {22.1} \\
0761670701 & 0.1 & 0.001 & 2.43$\pm$0.24&1.82$\pm$0.88&3.41$\pm$0.26  & {~9.8}\\
0761670801 & 0.15 & 0.001 & 2.92$\pm$0.48&-&4.56$\pm$0.48  &13.3\\
0761670901 & 0.11 & 0.001 & 2.76$\pm$0.24&-&4.14$\pm$0.27  &17.5\\
\hline 
\end{tabular}
\end{table*}

\begin{subfigures}
\begin{figure*}[t]
\centering
%\vspace*{-0.2in}
\vspace*{0.1in}\epsfig{figure= 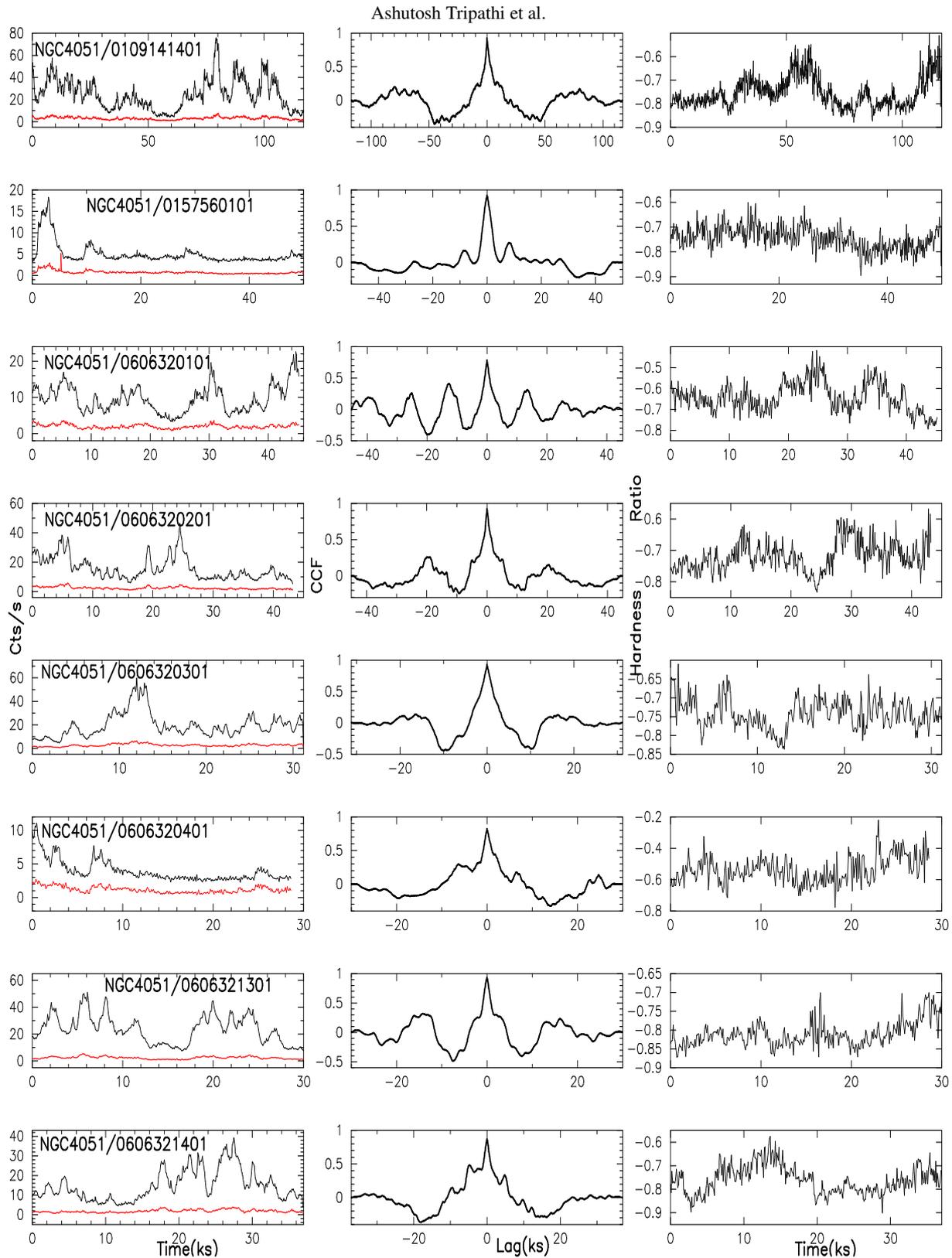,height=8.0in,width=6.0in,angle=0}
\caption{Soft (0.3--2 keV, in black) and hard (2--10.0 keV, in red) light curves (left panels).
Cross correlation functions (CCFs) between those bands (middle panels) and 
hardness ratios (right panels) are plotted for the observations of NGC 4051 labeled in the LCs.}
\label{}
\end{figure*}

\begin{figure*}
\centering
%\vspace*{-0.2in}
\vspace*{0.2in}\epsfig{figure= 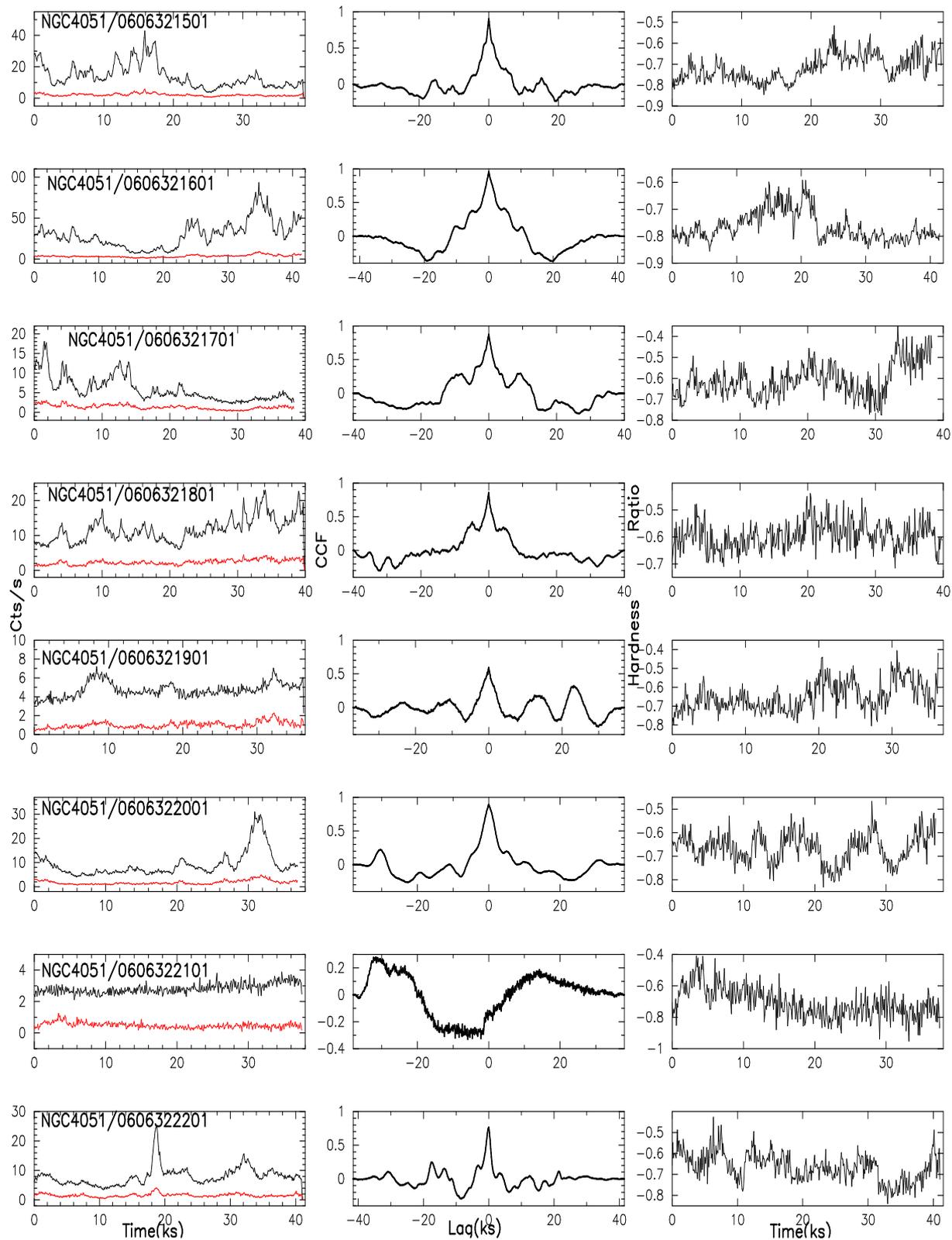,height=8.0in,width=6.0in,angle=0}
\caption{As in Fig.\ 3a.}
\label{}
\end{figure*}

\begin{figure*}
\centering
%\vspace*{-0.2in
\vspace*{0.2in}\epsfig{figure= 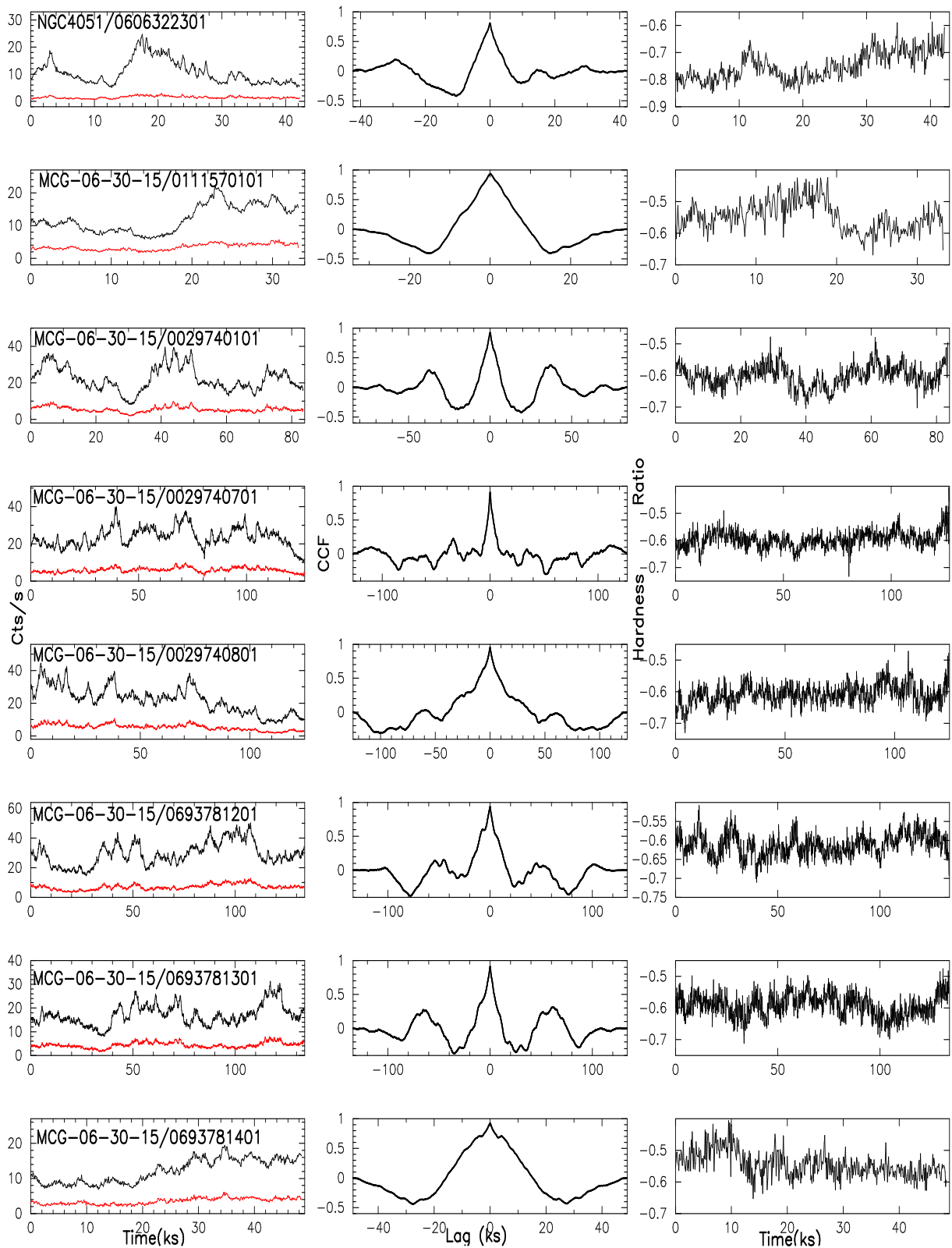,height=8.0in,width=6.0in,angle=0}
\caption{As in Fig.\ 3a for NGC4051 and MCG-06-30-15.}
\label{}
\end{figure*}

\begin{figure*}
\centering
%\vspace*{-0.2in}
\vspace*{0.2in}\epsfig{figure= 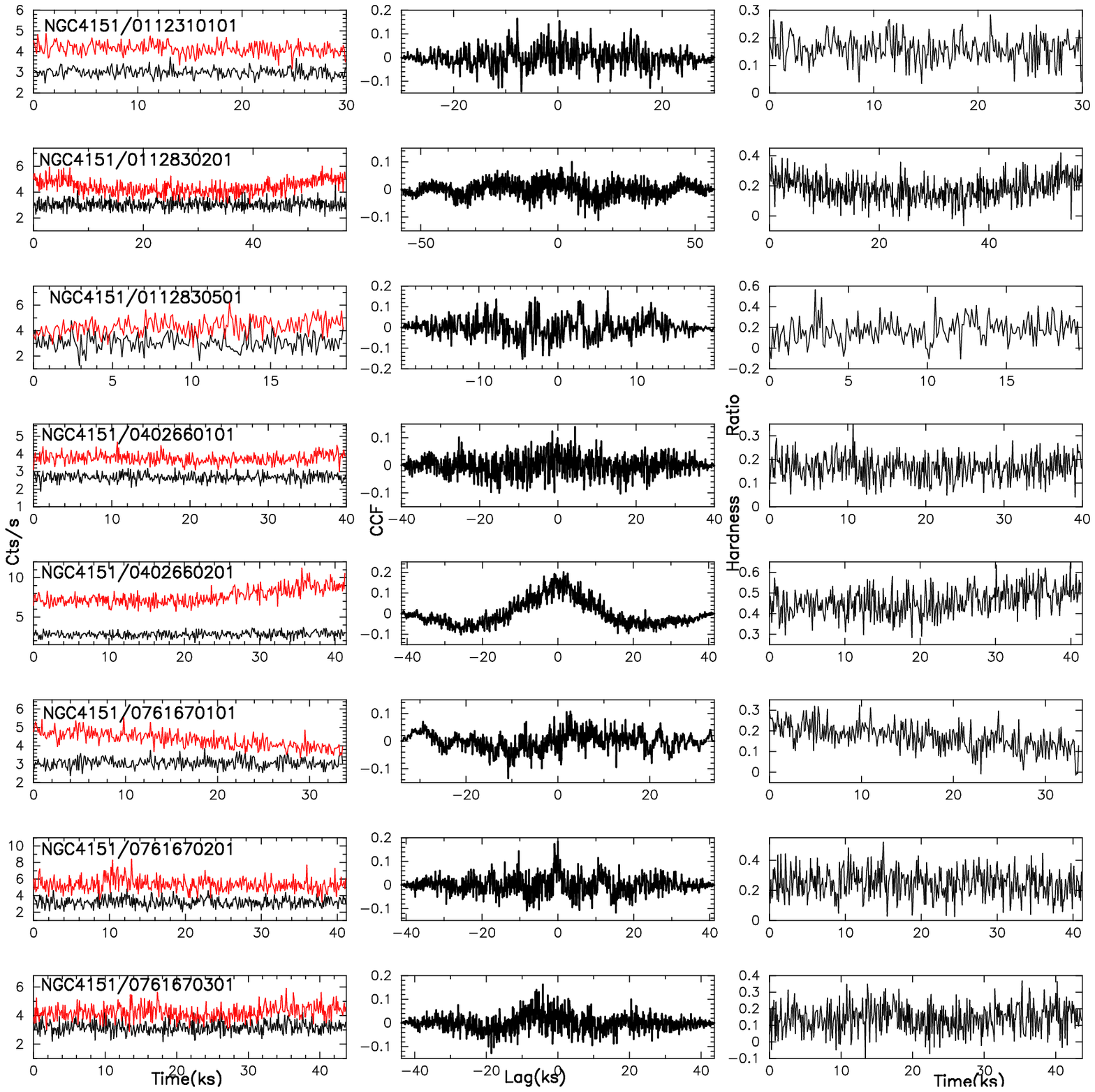,height=8.0in,width=6.0in,angle=0}
\caption{As in Fig.\ 3a for NGC4151.}
\label{}
\end{figure*}

\begin{figure*}
\centering
%\vspace*{0.2in}
\vspace*{0.2in}\epsfig{figure= 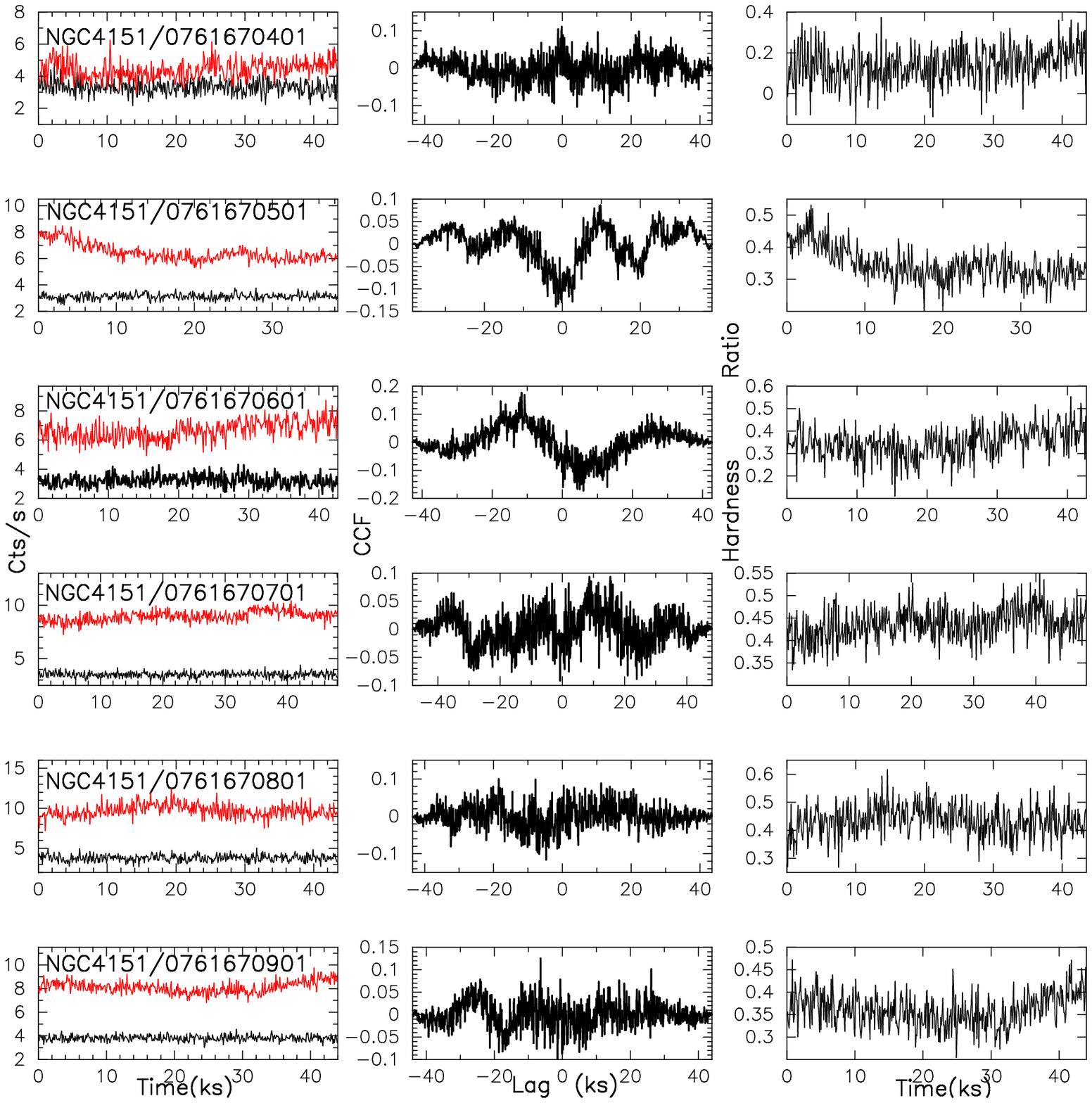,height=6.0in,width=6.0in,angle=0}
\caption{As in Fig.\ 3a for NGC4151.}
\label{}
\end{figure*}

\end{subfigures}

A tight correlation between extreme 
ultraviolet emission from {\it EUVE} and continuum X-ray emission from {\it RXTE} was found  by Cagnoni et 
al.\ (1999) who concluded that its SMBH  mass is $2.0-8.0\times10^7 M_{\odot}$.
{\it Chandra} observations of NGC 4051 in 2000 (Collinge et al.\ 2001) 
indicated the absence of significant soft X-ray emission.

Breedt et al.\ (2010) presented a cross-correlation study using X-ray and optical light curves spanning more than
12 years which  confirmed that the amplitude of optical variability is much
 less than that of X-rays. Light curves in the optical band were found to be well correlated at around 1 day, which is
consistent with the time taken by light to cross the optical emitting region and suggests that optical variations 
are driven by X-ray reprocessing.
Alston et al.\ (2013) used {\it XMM-Newton} and {\it Swift} data to analyze light curves of NGC 4051 and found 
no significant 
correlation between UV and X-ray emission, though the  UV was variable on both short and long time-scales.

Out of the 17 {\it XMM-Newton} observations of NGC 4051, one was taken in
2001, one in 2002 and the rest in 2009. We note that the mean count rates for the full band show 
the source was in a high flux state during the 2001 observation while it was in low flux state in 2002 observation. 
We also note that these mean count rates for the full band varied by a factor of 10 over the course of 3 weeks 
in 2009, and that they frequently change by a factor of 2 or more between subsequent observations that are only 
2 or 4 days apart. In every case for these observations of NGC 4051, the ACFs gave putative timescales where 
minima are seen and where the monotonic declines away from the peak at $t = 0$ are followed by 
%(usually modest) 
rises (Pandey et al.\ 2017); these dip timescales are listed as $\tau$ in Table 2 and range from 2.9 ks 
through 45.3 ks.
The mean value of these timescales 
$11.16\pm9.91$ ks. In no case were there repetitive dips at nearly the same multiples of a lag time that might 
signal a QPO.

We see that for every observation but one, the CCFs peak at zero lag and have correlation coefficients 
exceeding 0.8 at that time. This implies that the hard and soft fluxes typically vary synchronously and their 
photons arise from essentially the same physical process and are apparently co-spatial.  The exception to this 
strong correlation is seen in the CCF computed for Obs ID 0606322101, but that data is very noisy, having the 
lowest count rates of all 17 observations, so the nominal lag of $\sim 17$ ks indicated by its CCF is not at 
all trustworthy. Using {\it Suzaku} data Miller et al.\ (2010) did find a time lag between its hard and soft 
bands of $\sim970$s while McHardy et al.\ (2004) noted the possibility of one of $\sim 3000$s in the 
{\it XMM-Newton} data for 2001.
  
We note from the separate LCs for the soft and hard X-rays in the left columns of those figures that the soft 
flux tends to vary proportionately more than does the hard flux.  That translates into the HRs usually getting 
softer (more negative) during flares and harder  during quiescent periods, as is clear from a comparison of the LCs 
and the HRs for each source.  There are, however, exceptions to that trend. One such is for the single observation 
made during 2002 (Obs.\ ID 0157560101) when the HR remained essentially flat, despite the presence of one major 
flare and a couple of minor ones.  The other three exceptions, however, correspond to the faintest states and 
so may be dominated by noise. 

\subsection{MCG$-06-30-15$}

MCG$-06-30-15$ is a NLSy1 galaxy situated at a distance of 37 Mpc ($z =$ 0.008). It was first observed as a point 
X-ray source in June 1976 with the \textit{SAS-3} X-ray satellite (Pineda et al.\ 1979). 
Vaughan et al.\ (2003) reported a strong correlation between rms variability amplitude and flux which was previously seen only in X-ray binaries.  Arevalo et al.\ (2006) analyzed the 300 ks 2001  {\it XMM-Newton} observation to study 
correlation between 0.2-10.0 X-ray and 300-400 nm  UV bands and found them to be significant. 

Of the 7 {\it XMM-Newton} observations of MCG-06-30-15, one was taken in
2000, three in 2001 and the other three in 2013. We note that these mean count rates for the full band show 
the source was in moderate and high flux states during the 2001 and 2013 observations while it was in a somewhat lower flux state 
during the 2000 observation. We also note that these mean count rates for the full band are almost the same in 2001 and 2013. 
In every case for these observations of MCG$-06-30-15$, the ACFs gave putative timescales where minima are seen 
and where the monotonic declines away from the peak at $t = 0$ are followed by 
%(usually modest) 
rises (e.g.\ Pandey et al.\ 2017); these dip timescales are listed as $\tau$ in Table 2 and range from 10.6 ks through 44.0 ks. 
In no case were there repetitive dips at nearly the same multiples of a lag time that might signal a QPO.

The separate LCs for the soft and hard X-rays in the left columns of those figures show that the soft 
flux tends to vary proportionately more than does the hard flux. Thus, the HRs usually are
softer during flares and harder  during quiescent periods. 

\subsection{NGC 4151}

NGC 4151 is a close-by ($z=0.003319$) and well studied Seyfert 1.5 galaxy possessing a clear
bright stellar nucleus. 
It was observed by the \textit{UHURU} satellite during 1970-1971 (Gursky et al.\ 1971) when it was  detected and reported as an X-ray source for the first time. 

Oknyanskij et al.\ (2006) studied correlations between the infrared and optical variability and found that the time delay between optical and NIR variation 
changes with time and is correlated with the state of nuclear processes in the source. Another correlation between the
optical luminosity of this AGN and the time lag between UV/optical and NIR light curves was interpreted as thermal dust 
reverberation (Minezaki et al.\ 2006). 
Onken et al.\ (2014) found the SMBH mass of this source to be $3.76\pm 1.15 \times 10^7$ M$_{\odot}$. Keck et al.\ (2015) 
presented the X-ray timing and spectral analysis of simultaneous 150 ks of {\it NuSTAR} and {\it SUZAKU} observations taken during 
November 2012 and found strong evidence of relativistic reflection from inner accretion disk. Landt et al.\ (2015) 
studied coronal line variability in this source using simultaneous optical, IR and X-ray observations. Recently, Edelson 
et al.\ (2017) analyzed the {\it SWIFT} observation of the source taken in 2016 and reported UV/optical bands to be significantly correlated 
with X-rays. The variability within those UV/optical bands  also is correlated.

{\it XMM-Newton} provided a total of 14 LCs of NGC 4151 of which three were taken in
2000, two in 2006 and the remaining nine in 2015.  The mean count rates for the full band show the source 
was in moderate and low flux states during most of the observations. 
We also note that these mean count rates for the full band did not even vary by a factor of 2 over the entire set 
of observations. In every case for these observations of NGC 4151, the ACFs gave putative timescales where minima are seen 
and where the monotonic declines away from the peak at $t = 0$ are followed by 
%(usually modest) 
rises; these dip timescales are listed as $\tau$ in Table 2 and range from 9.8 ks through 30.8 ks. 
Like the other sources, in no case were there repetitive dips at nearly the same multiples of a lag time that might signal a  QPO.

 The CCFs for NGC 4151 are much weaker than for the other two sources, rarely exceeding 0.15.  These thus usually show no preference for any particular lag, or at most for a mild one  at 0 lag. This result can be attributed primarily to the weaker variations exhibited by this source that make computations of lags difficult and secondarily to the somewhat lower overall count rates.  The spectra of NGC 4151 are substantially harder than those of the other two sources, having positive instead of negative HR values.  They show a mild tendency to be harder when brighter, though there are no strong flares seen during any of the 14 observations.  
 
\begin{figure}
\centering
\vspace*{0.2in}
\epsfig{figure= 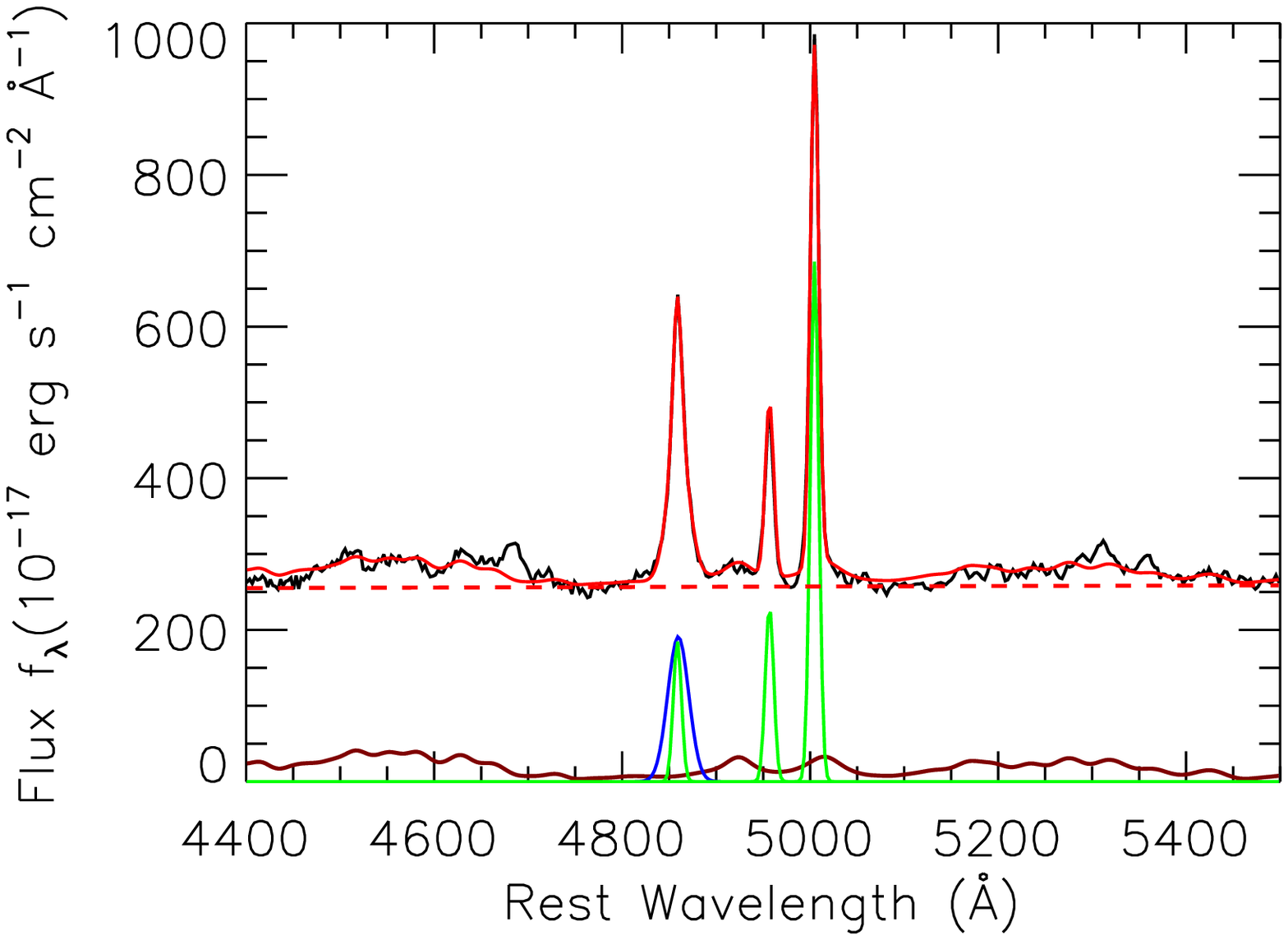,height=2.8in,width=2.8in,angle=0}
\caption{Fit to the spectrum of NGC 4051. The continuum is modelled with a power law (red dashed line) and Fe II emission (brown line). The $H\beta$ line (4861 \AA) is fitted with one gaussian for the
broad component (blue) and one gaussian (green) for the narrow component.
  }
\label{bhmassngc4051}
\end{figure}

\begin{figure}
\centering
\vspace*{0.2in}
\epsfig{figure= 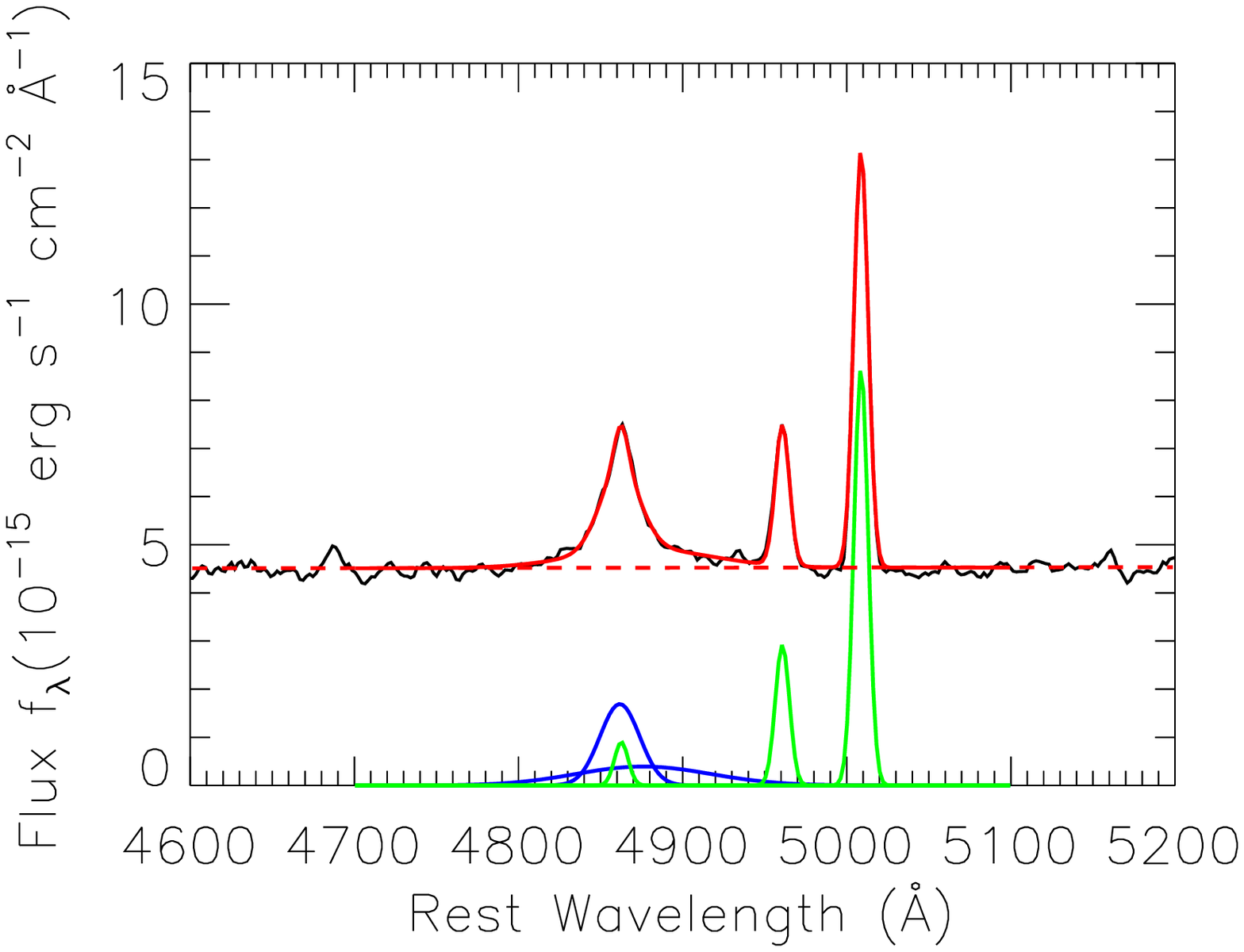,height=2.8in,width=2.8in,angle=0}
\caption{Fit to the spectrum of MCG$-06-30-15$, as in Fig.\ 5.   }
\label{bhmassngc4051}
\end{figure}

\begin{figure}
\centering
\vspace*{0.2in}
\epsfig{figure= 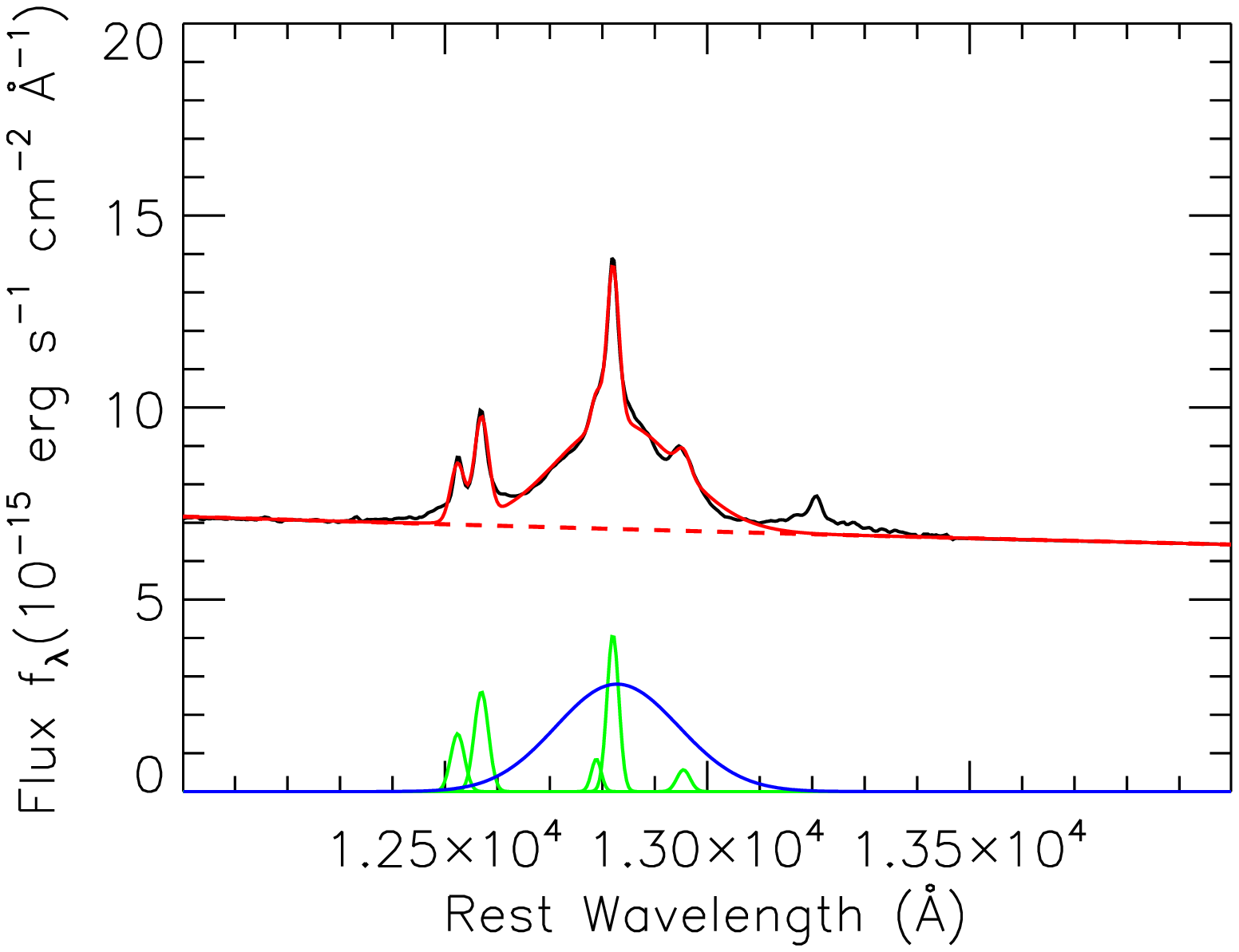,height=2.8in,width=2.8in,angle=0}
\caption{Fit to the spectrum of NGC 4151. The continuum is again modelled with a power law (red dashed line). The Pa$\beta$ line (12820 \AA) is fitted with one gaussian for the
broad component (blue) and a gaussian (green) for the narrow component.  
  }
\label{bhmassngc4151}
\end{figure}

\section{Black hole mass estimation}

In order to study the difference of BH masses estimated in a uniform way between these NLSy1 and Seyfert 1.5 galaxies, we analysed the optical or near-infrared (NIR) spectrum for these three sources. The local continuum of the  line we focussed on (H$\beta$ for NGC 4051 and MCG-06-30-15, and Pa$\beta$ for NGC 4151) is modelled with single power law (plus iron emission, if necessary). We used Gaussian functions to fit the emission line after subtracting the continuum (Guo \& Gu 2014, Shen et al.\ 2011).

The nuclear optical spectroscopy of NGC 4051 is taken from the NED\footnote{http://ned.ipac.caltech.edu/} database, where the data were taken at the  2.3m Bok Telescope (Moustakas \& Kennicutt 2006 ). For this source (Fig.\ 4), H$\beta$ can be well modelled by one broad component with full width at half maximum (FWHM) at 1625.4 km s$^{-1}$ and a narrow component. The black hole (BH) mass is calculated by using this FWHM and the luminosity of $H\beta$ line (log ($L_{H\beta}$) = 38.81 erg s$^{-1}$; see Vestergaard \& Peterson 2006). This gives us $M_{BH} = 0.12(\pm 0.11 ) \times 10^6 M_{\odot}$.  The result is found to be consistent,  within the errors, with the value found using reverberation mapping in Denney et al.\ (2010).
 
The optical spectral data of MCG$-06-30-15$ were obtained from an observation in 2008 in the public archive of the Lick Observatory 3m Shane telescope (Bentz et al.\ 2009).
As Fig.\ 5 demonstrates, the $H\beta$ line is characterized by two broad components with FWHM at 1993 km s$^{-1}$ and one narrow component. 
The BH mass is calculated in the same way as for NGC 4051 (log ($L_{H\beta}$) = 40.09) and found to be $M_{BH} = 1.15(\pm 0.8) \times 10^6 M_{\odot}$. This black hole mass estimation is in reasonable agreement with that found in Bentz et al.\ (2016). 

The near-infrared spectrum data of the spectra of NGC 4151 was also download from the NED database. It was observed by the NASA 3m Infrared Telescope Facility (IRTF) on 2002 April 23 and was discussed by Riffel et al.\ (2006). For NGC 4151 (Fig.\ 6), we use one Gaussian to fit the broad component of  the Pa$\beta$ line with FWHM of 6445.2 km s$^{-1}$ and one Gaussian for the narrow component. The SMBH mass is estimated by using the broad FWHM and luminosity of Pa$\beta$ (log ($L_{\rm{Pa}\beta}$) = 40.27 erg s$^{-1}$; see Kim et al.\ 2010), which gives the estimate for the mass of the black hole to be $M_{BH} = 8.2(\pm 0.5 {\rm ~dex}) \times 10^7 M_{\odot}$.  Our value  is somewhat higher than that found in Bentz et al.\ (2006; $4.57\times 10^7 M_{\odot}$) but is consistent considering the uncertainties.  

\section{Discussion and Conclusions}

We have examined the LCs for timescales using ACFs and have also considered the cross-correlations between the hard and soft bands and hardness ratios. NGC 4051 was quite active during essentially all of these observations, with individual LCs often showing flares that produce a change of a factor of 5 or more in count rates.  Over the entire set of observations the mean X-ray fluxes (0.3--10 keV) of the individual observations varied by just about a factor of 10.  The more limited observations of MCG$-06-30-15$ showed a great deal of activity during individual stares, with fluxes usually varying by factors of 2--3, but relative stability overall, with a mean counts changing by only about a factor of 2.  While there were also many measurements made of NCG 4151, this source was not doing much of interest, typically changing by less than 20\% during an individual LC, though the overall flux range was roughly a factor of 2.

In the case of NGC4151, the ACF values always dropped down to much lower values very soon after zero lags as compared to other sources; this implies  that the strength of the signal relative to the noise is low.  For the other two objects the ACF values almost always fall more gradually, indicating that  the strength of signal is high and the noise does not dominate.  Hence the variability timescales obtained for NGC 4051 and MCG$-6-30-15$ are more reliable as the noise contribution is low.

Our cross-correlations between the hard (2--10 keV) and soft (0.3--2 keV) bands show that they are very well correlated on sub-day timescales for NGC 4051 and MCG-6-30-15. Using the hardness ratio to get a feel for the spectral variability, we find that these two sources typically soften during flares.  

The independent estimates of the black hole mass estimates we have made are in reasonable accord with earlier measurements.  They reinforce the hypothesis that NLSy1 galaxies have relatively low mass black holes and hence, high Eddington ratios. 

\section*{ACKNOWLEDGMENTS}
We thank the anonymous referee for useful comments and suggestions which helped us to improve the manuscript.
This research is based on observations obtained with {\it XMM-Newton}, an ESA science mission with instruments and 
contributions directly funded by ESA Member States and NASA. This research has made use of the NASA/IPAC Extragalactic Database (NED) which is operated by the Jet Propulsion Laboratory, California Institute of Technology, under contract with the National Aeronautics and Space Administration.

AT acknowledges support from the China Scholarship Council (CSC), grant no.\  2016GXZR89.  PJW is grateful for hospitality at SHAO while this paper was written.  ACG was partially supported by the Chinese Academy of Sciences (CAS) President's International Fellowship 
Initiative (PIFI), grant 2016VMB073.
MFG is supported by the National Science Foundation of China (grants 11473054 and U1531245) and by the Science 
and Technology Commission of Shanghai Municipality (grant 14ZR1447100).

\clearpage

\end{document}